\documentclass[english,a4paper,12pt]{article}
\usepackage[T2A]{fontenc}
\usepackage[cp1251]{inputenc}
\usepackage{amsmath}
\usepackage{graphicx}
\usepackage{amssymb}
\usepackage{epsf}
\usepackage{pazh}
\usepackage[labelsep=period,hang]{caption} 

\tightenlines

\newcommand{\WI}[2]{#1_{\mathrm{#2}}}
\sloppypar

\begin{document}
\baselineskip 21pt


\title{\bf Low-Mass Neutron Stars with Rotation}

\author{\bf 
A.V. Yudin\affilmark{1,2*}, T.L. Razinkova\affilmark{1}, S.I. Blinnikov\affilmark{1,3}}

\affil{ $^1${\it Institute for Theoretical and Experimental Physics,
ul. Bolshaya Cheremushkinskaya 25, Moscow, 117259 Russia}\\
$^2${\it ``Kurchatov Institut'' National Research Center, pl. Kurchatova 1, Moscow, 123182 Russia}\\
$^3${\it Kavli IPMU, Tokyo University, Kashiwa (WPI), Japan}}

\vspace{2mm}

\sloppypar \vspace{2mm} \noindent Abstract --- The properties of low-mass neutron stars with rigid rotation are considered. The possible
evolution paths of such stars in a close binary system with mass transfer are calculated. The properties
of the gamma-ray burst GRB 170817A interpreted in terms of the stripping model, a natural ingredient of
which is the explosion of a low-mass neutron star --- a binary component, are briefly discussed.



\vfill
\noindent\rule{8cm}{1pt}\\
{$^*$ email: $<$yudin@itep.ru$>$}

\clearpage

\section*{INTRODUCTION}
\noindent Neutron stars (NSs), along with black holes,
are the final point of evolution of massive stars
exploding as supernovae. They possess surprising,
as yet incompletely clear properties, while the matter
in their interiors is in the most extreme state of
all those available in the present-day Universe. It
is well known that NSs have a maximum mass
lying in the range  $2\leq M/M_\odot\leq 3$  (Rhoades and
Ruffini 1974; Demorest et al. 2010). High-mass
NSs get rapt attention from both observers and
theoreticians, because the exact maximum mass
depends critically on the unknown equation of state
of matter with a density exceeding the nuclear one
(Haensel et al. 2007). However, NSs also have
a minimum mass of $0.1 M_\odot$ (see, e.g., Haensel
et al. 2002). NSs of such low masses attract less
attention, first, because their properties are known
comparatively better: the matter inside them is in a
much less extreme state than in massive NSs due to
their lower density. Second, it is still unclear whether
a NS of such a low mass can be obtained from the
core collapse of a massive star and the subsequent
supernova explosion. Apparently, this cannot be done
directly. A NS can reach a minimum mass only as
a result of its evolution in a close binary system of
neutron stars.

Clark and Eardley (1977) were among the first
who proposed and considered in detail this scenario.
In their calculations two stars with masses of $1.3 M_\odot$
and $0.8 M_\odot$ approached each other due to the loss of
angular momentum by the binary through the radiation
of gravitational waves, and mass transfer from
the less massive, but larger (in size) companion to
the more massive one began at some instant. Having
reached the minimum NS mass, the low-mass
neutron star (LMNS) lost its hydrodynamic stability
and exploded. This process was subsequently proposed
by Blinnikov et al. (1984) as a source of short
gamma-ray bursts (GRBs). In the succeeding paper
by Blinnikov et al. (1990) D.K. Nadyozhin carried out
one-dimensional hydrodynamic spherically symmetric
calculations of the explosive disruption of a LMNS
that reached its minimum mass. Surprisingly, the
kinetic energy of the explosion turned out to be very
close to the classical energy of a supernova explosion
$\sim 1$~Bethe, i.e., $10^{51}$~erg. This led Imshennik (1992)
to the formulation of a rotational supernova explosion
mechanism, where a close NS binary results from the
collapse and subsequent fragmentation of the core
of a massive star. The evolution of such a NS binary,
which ends with the explosion of its low-mass
component, ultimately leads to the explosion of the
entire star as a supernova. The LMNS explosion
was considered by Colpi et al. (1989) and Sumiyoshi
et al. (1998), who, among other things, studied the
properties of the burst of neutrino radiation accompanying
the explosion and the nucleosynthesis processes.
Having taken the parameters from Blinnikov
et al. (1990) as initial data, Manukovskiy (2010) carried
out three-dimensional self-consistent hydrodynamic
calculations of the explosion of a LMNS in the
orbit of a massive companion. Many historical details
of the development of this scenario can be found in the
review by Baklanov et al. (2016).

However, the interest in this mechanism of short
GRBs decreased with time, because their observed
energies were, as a rule, much greater and their
spectra were harder than those predicted by Blinnikov
et al. (1990) (see, e.g., Hamidani et al. (2019)
and references therein). GRB 170817A associated
with the signal GW170817 at the LIGO and
Virgo gravitational--wave interferometers (Abbott
et al. 2017a) helped to revive the interest. This
GRB turned out to be peculiar: with a low isotropic
energy (Abbott et al. 2017b), spectral peculiarities
(Villar et al. 2017), and a large estimated ejecta mass
(Siegel 2019). In addition, no traces of the presence of
a strong jet were observed either (Dobie et al. 2018).
All these characteristic features of GRB 170817A are
difficult to explain in terms of the standard NS merger
scenario, where two NSs merge into one object --- a
supermassive NS or a black hole. On the other hand,
these features are naturally explained in terms of the
Blinnikov et al. (1984) mechanism, which we will call
below the stripping model.

A detailed discussion of the merger and stripping
models and their comparison with observational
data will be given in an appropriate place (Blinnikov
et al. 2020), while here we will only note that the
realizability of these scenarios is determined mainly
by the initial binary mass ratio (asymmetry). A nearly
vertical segment corresponds to NSs with a mass of
the order of the solar one on the NS mass–-radius diagram
(see, e.g., Lattimer and Prakash 2001). Thus,
the radius of such NSs depends weakly on their mass
(an effective polytropic index $n \approx 1$), and during their
merger they behave like two liquid droplets and merge
into one object. Note that almost all of the calculations
of this process performed to date dealt precisely
with the case of equal and fairly large masses. Even
in the case where a large mass ratio was considered
(Dietrich et al. 2017), the mass of the less massive
component was fairly large (of the order of the solar
one). If, however, the mass ratio is great and the
LMNS mass is fairly small, then the stripping scenario
should be realized. The specific NS mass, small
enough for the onset of stripping, depends on the
equation of state of matter in the nuclear and, what
is more important in this case, subnuclear region. As
shown by Sotani et al. (2014), there are significant
uncertainties in the behavior of the LMNS mass--
radius curves. However, an analysis of fig.~1 from
this paper allows the characteristic value of this mass
to be estimated as $M\sim 0.5 M_\odot$. We will call NSs
with masses smaller than this value low-mass ones
(LMNSs).

\section*{FORMULATION OF THE PROBLEM}
\noindent In most of the studies cited above the binary NS
spin was neglected. To a first approximation, this is
completely justified for a dense massive component.
However, a LMNS has a peculiar structure: a tiny
dense compact core containing the bulk of the mass
and an extended envelope. The rotation effects may
turn out to be important for such a structure.

We will consider the influence of the LMNS spin
on its parameters and the possible evolution paths. In
this case, we will neglect the tidal deformation from
the massive companion by assuming the distribution
of matter in the LMNS to be axisymmetric. The
validity of this approximation breaks down only for
the outer, least dense layers whose structure is determined
by the isolines of a modified (with the LMNS
spin) Roche potential. Determining the LMNS shape
generally requires solving a three-dimensional problem
in this case.

In addition, we will deem the LMNS rotation law
to be the rigid one, i.e., will assume that the characteristic
angular momentum redistribution time inside
the NS is much shorter than the mass transfer time.
This assumption can break down only at the final
evolutionary stages of a NS binary system.

Since the original \textsc{ROTAT} code (for its description,
see below) used by us for the calculation of rotating
configurations is Newtonian, i.e., it disregards the
general relativity effects, we will restrict our study to a
LMNS with a mass $M\leq 0.2 M_\odot$. The radius of such
stars is $R\geq 20$~km, and the relativistic parameter
turns out to be fairly small: $\frac{2 G M}{R c^2}\leq 0.03$.

\subsection*{The Equation of State}
\noindent We will use the fits for the dependence of pressure
on density $P=P(\rho)$ proposed by Haensel and
Potekhin (2004) as the equations of state (EOS)
of matter. They describe the properties of NS
matter at temperature $T=0$ in a wide range of
densities both in the subnuclear region and at densities
above the nuclear one. The convenient subroutines
written in \textrm{FORTRAN} that compute various
thermodynamic quantities for several possible
EOS are presented at the site of one of the authors
(http://www.ioffe.ru/astro/NSG/BSk/). At
low densities ($\rho\lesssim 3\times 10^{5}~\mbox{g}\cdot\mbox{cm}^{-3}$) these fits are
inaccurate, and we smoothly join them with the
$n=3/2$ polytrope ($P=P_0\rho^{5/3}$). The parameters of
non-rotating LMNSs calculated using some of these
EOS are given in Table 1. 
\begin{table}[t]
\vspace{6mm}
\centering
{{\bf Table 1.} Properties of minimum-mass NSs for various
equations of state}

\vspace{5mm}\begin{tabular}{|c|c|c|c|} \hline\hline
EoS & $\WI{M}{min}$ [$M_\odot$] & $R$ [km] & $\WI{\rho}{c}$ [$\mbox{g}\cdot\mbox{cm}^{-3}$] \\
\hline
BSk19   &   $0.097$     &   $208$   &      $1.8\cdot  10^{14}$        \\
BSk22   &   $0.089$     &   $272$    &   $1.5\cdot  10^{14}$          \\
BSk24   &    $0.093$    &   $238 $  &      $1.9\cdot  10^{14}$       \\
BSk25   &   $0.091 $    &    $233$         &     $2.3\cdot  10^{14}$        \\
BSk26   &   $0.096 $    &    $222$         &     $1.8\cdot  10^{14}$        \\
\hline
\end{tabular}
\label{TableEoS}
\end{table}
The first, second, third,
and last columns provide the EOS abbreviation,
the corresponding minimum mass, the radius, and
the central density, respectively. Everywhere below
we will use the BSk19 fit as the basic case for our
calculations.

\subsection*{The Rotat Code}
\noindent Let us briefly describe the algorithm implemented
in the \textsc{ROTAT} code (Aksenov and Blinnikov 1994).
Consider an axisymmetric stellar configuration in a
state of stationary rotation. The equation of motion
of a fluid element is
\begin{eqnarray}
   {  \rho (\mathbf{v}{\nabla} )\mathbf{v}
    +{\nabla} P+\rho{\nabla}\Phi}=0, \label{eq:stat0}
\end{eqnarray}
where {$\mathbf{v}$} is the velocity, ${\rho}$ is the matter density, ${\Phi}$ is
the gravitational potential that satisfies the Poisson equation
\begin{equation}
    {\nabla^2 \Phi = 4\pi G\rho} ,
    \label{eq:Poisson}
\end{equation}
and ${P}$  is the matter pressure related to the density by
the barotropic equation of state
\begin{equation}
    { P=P(\rho)}.
    \label{eq:bareos}
\end{equation}
For the barotropic equation of state the surfaces
of constant pressure and constant density coincide,
while the angular velocity of rotation $\omega(\xi)$ depends
exclusively on distance $\xi$ from the rotation axis, i.e.,
it is constant along the cylinders coaxial with the
rotation axis (the Poincare theorem, see Appendix A).
Thus, the linear velocity of the fluid element is related
to its angular velocity by the expression
\begin{equation}
    { \mathbf{v}
    =\xi\omega(\xi)\mathbf{e}_{\varphi}} ,
\end{equation}
where $\mathbf{e}_{\varphi}$ is a unit vector in the direction of the azimuthal
angle $\varphi$. Integrating the equation of motion (\ref{eq:stat0}), we can obtain the Bernoulli integral
\begin{equation}
    H(\rho)+\Phi+\Psi=C,
    \label{eq:stat1}
\end{equation}
where $C$ is the constant of integration, $H$ is the
enthalpy defined by the expression
\begin{equation}
     {H(\rho)=\int\limits^{P(\rho)}\frac{{\rm d} P'}{\rho'}},
     \label{eq:enth}
\end{equation}
and $\Psi$ is the centrifugal potential related to the angular
velocity by the expression
\begin{equation}
     {\Psi=-\int\limits^\xi \omega^2(\xi') \xi'{\rm d}\xi'}.
\end{equation}
For rigid rotation
\begin{equation}
 \Psi=-\frac{C_\psi}{2} \xi^2.
\end{equation}

If we use the Bernoulli integral (\ref{eq:stat1}) to express the
density as a function inverse to the enthalpy and
substitute this into the Poisson equation (\ref{eq:Poisson}), then
we will obtain an equation containing the gravitational
potential and the constants $C$ and $C_\psi$. This
equation is then written in finite differences on a two--dimensional
grid $(r,\vartheta)$, where $r$ is the distance from
the configuration center and $\vartheta$ is the polar angle measured
from the rotation axis. The radial coordinate on
the grid outside the sphere enclosing the entire star is
a quantity inversely proportional to the distance to the
center (Clement 1974).

The boundary conditions correspond to symmetry
at the center: for $r\rightarrow 0$ we have
\begin{equation}
\frac{\partial\Phi}{\partial r}\bigg|_{r=0}\!\!=0, \quad \frac{\partial^2\Phi}{\partial r^2}\bigg|_{r=0}\!\!=\frac{4\pi}{3}G\WI{\rho}{c},
\end{equation}
where $\WI{\rho}{c}$ is the density at the stellar center (the erratum
in the original paper by Aksenov and Blinnikov
(1994) in the grid form of the central condition
is corrected in Appendix B). The condition $\Phi \rightarrow 0$ is
specified for $r \rightarrow \infty$. In addition, the ratio of the polar
and equatorial radii $\theta = \WI{R}{p} / \WI{R}{e}$ and the maximum
density allow $C$  and $C_\psi$ to be determined simultaneously
with the field $\Phi$.

The density at each grid point can be found using
(\ref{eq:stat1}). The position of the grid point at which the
maximum density is reached can be found during
the calculations. Owing to the use of the differential
relation (\ref{eq:Poisson}) between the gravitational potential
and density instead of the integral expression for the
potential via the density, the matrix of coefficients
for the system of difference equations is sparse, i.e.,
there are many zero elements in it. The system of
difference equations was solved by the quadratically
converging Newton method. In this case, we used
the method of solving a system of linear equations
(obtained by linearizing the complete system in the
Newton method) with sparse matrices of coefficients
described in the book by Osterby and Zlatev (1983).
The accuracy of our calculations was checked by the
virial test
\begin{equation}
VT = \frac{1}{|W|}\Big|2T + W + 3\!\int\! P dV\Big| ,
\end{equation}
where $T$ is the rotational kinetic energy of the star and
$W$ is its gravitational energy.

\section*{RESULTS OF OUR CALCULATIONS}
\begin{figure}[htb]
\epsfxsize=16cm \hspace{-2cm}\center\epsffile{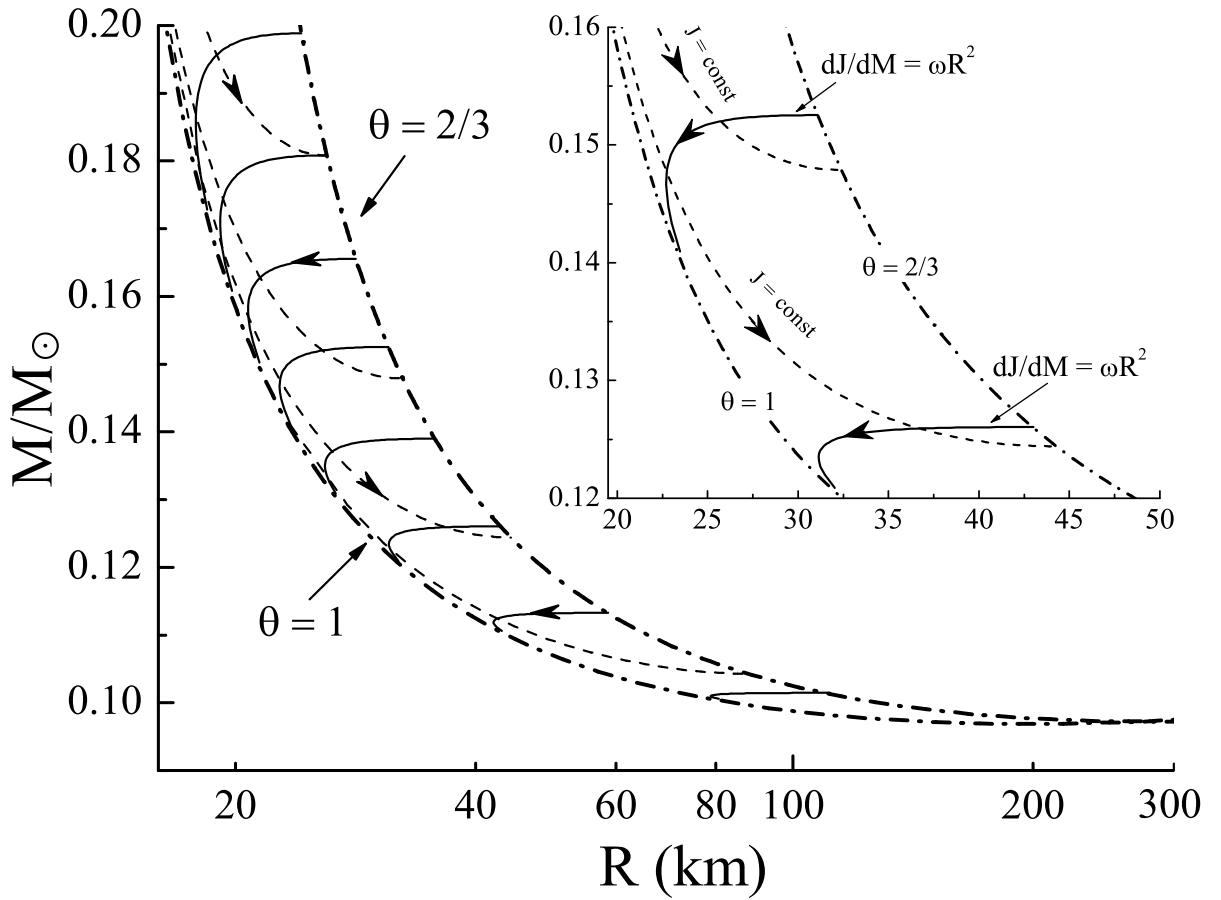}
\caption{\rm LMNS mass–radius diagram. The dash–dotted lines bounding the horn-shaped region correspond to the sequence of
nonrotating configurations with $\theta=1$ and configurations with fixed oblateness $\theta=2/3$. The dashed lines give the sequence
of stars with a constant total angular momentum $J$. The solid lines correspond to the law of evolution $dJ/dM=\omega R^2$. For the
remaining details, see the text.} \label{Pix-Main}
\end{figure}

\noindent The main results of our calculations are presented
in Fig.~\ref{Pix-Main}. It shows the lower part of the mass--radius
$(M{-}R)$ diagram for LMNSs. In the inset the central
part of the figure is shown on an enlarged scale. The
dash–dotted lines bounding the narrow horn-shape
region correspond to configurations with fixed oblateness
$\theta\equiv\WI{R}{p}/\WI{R}{e}$, where $\WI{R}{p}$ and $\WI{R}{e}$ are the polar and
equatorial radii, respectively. Everywhere below the
parameter $R$ on the diagram is equal to the equatorial
radius $\WI{R}{e}$ of the star. Naturally, $\theta=1$ corresponds to
non-rotating stars. On the other side the region of our
configurations is bounded by (approximate) $\theta=2/3$.
We failed to obtain more oblate configurations with
rigid rotation. The reason is explained in detail in
Appendix A. Here we will only say that at these values
of $\theta$ a mass outflow from the equator begins in rigidly
rotating LMNSs. Thus, all of the admissible configurations
are contained in the narrow band between
the $\theta=1$ and $\theta=2/3$ curves. These results are also in
complete agreement with the conclusions by Haensel
et al. (2002).

Let us now consider the possible evolution paths
of a LMNS in a binary system. The dashed lines
indicate the configurations with a constant total angular
momentum $(J=\mathrm{const})$. If the star could lose
its matter without any change in $J$ (for example, in
the form of a circumpolar outflow or a jet), then its
evolution on the $(M{-}R)$ diagram would be described
by one of the dashed curves in the direction indicated
by the arrows. Each such curve begins almost tangentially
from the $\theta=1$ line and gradually approaches
the equatorial outflow limit $\theta=2/3$. However, this
evolution path looks unrealistic.

Let us now turn to a more likely scenario: a LMNS
in a binary system fills its Roche lobe and begins to
transfer its mass to the more massive companion. In
this case, matter outflows virtually from the equator.
Depending on the orientation of the LMNS spin axis
relative to the orbital plane, the outflow point can
lie not exactly on the equator. Here we will restrict
ourselves to the simplest case of an equatorial outflow.
The star will then lose not only its mass, but also its
angular momentum. It is easy to show that this will
occur in accordance with the law
\begin{equation}
\frac{dJ}{dM}=\omega(\WI{R}{e}) \WI{R}{e}^2,
\end{equation}
where $\omega$ is the angular velocity of rotation, which is
constant over the star in the case of rigid rotation
considered by us, though, of course, it changes during
star’s motion across the $(M{-}R)$ diagram. The
corresponding trajectories are indicated by the solid
lines with arrows. As we see, the evolution path is
nontrivial in this case: initially the star moves leftward
across the mass–radius diagram, rapidly losing its
angular momentum, with the change in mass being
comparatively small. As the $\theta=1$ line of non-rotating
configurations is approached, the stellar track turns
and begins to asymptotically approach it, moving
downward and rightward. Thus, the LMNS evolution
in the binary system actually breaks up in this case
into two stages: at the first stage, it loses its angular
momentum and its radius decreases; at the second
stage, its evolution is virtually indistinguishable from
the evolution of non-rotating configurations: the radius
increases rapidly as the mass is lost.

\begin{figure}[htb]
\epsfxsize=16cm \hspace{-2cm}\center\epsffile{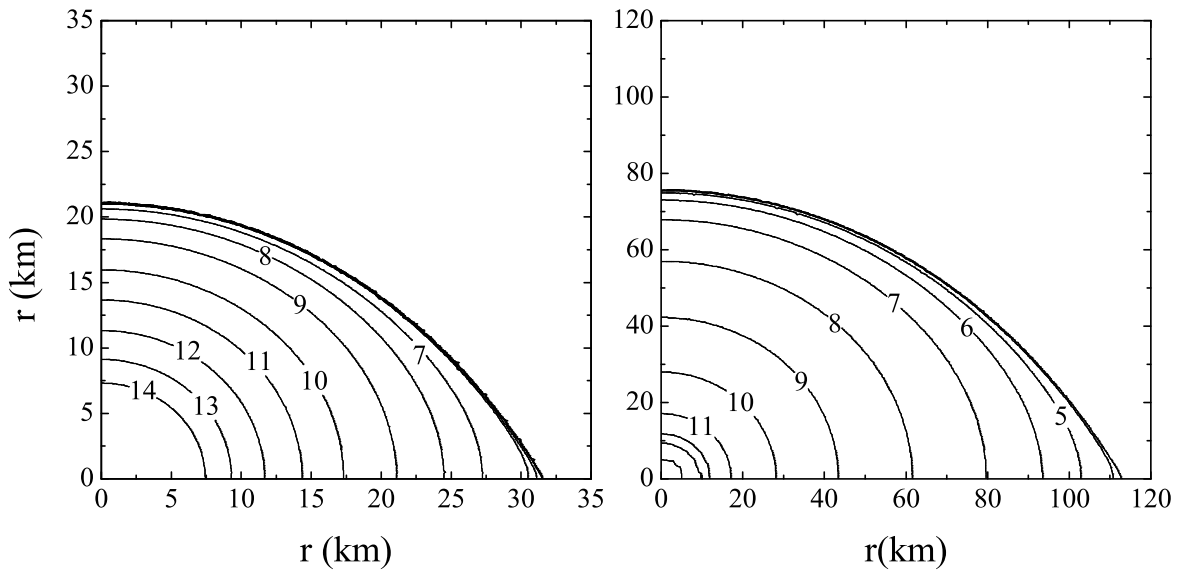}
\caption{\rm Spatial structure of critically ($\theta=2/3$) rotating stars with masses $M=0.15 M_\odot$ (left) and $M=0.1 M_\odot$ (right).
Density isolines are shown; the numbers correspond to $\lg\rho~\mbox{g}/\mbox{cm}^3$. The three central lines for the star with $M=0.1 M_\odot$
correspond to $\lg\rho = 12, 13\ \mbox{and}\ 14$; the corresponding numbers are not shown lest the graph be overloaded.} \label{Pix-Levels}
\end{figure}
Let us discuss two more questions. First, consider
the structure of a LMNS with critical rotation.
Figure~\ref{Pix-Levels} shows the structure of LMNSs with oblateness
$\theta=2/3$ and masses $M=0.15 M_\odot$ (left) and
$M=0.1 M_\odot$ (right). The curves with numbers are
isolines of the logarithm of density $\lg\rho~(\mbox{g}/\mbox{cm}^3)$.
On the right panel the central lines correspond to
$\lg\rho=12, 13, 14$; the corresponding numbers are not
plotted lest the figure be overloaded. As we see, the
star consists of a tiny dense nearly spherical core and
an extended deformed envelope (cf. the discussion in
Haensel et al. (2002)). In the case of $M=0.1 M_\odot$,
the contrast between the size of the core containing
the bulk of the mass and the envelope size is particularly
striking.

The configurations with critical rotation were
checked for stability in accordance with the criterion
developed in the paper by Bisnovatyi-Kogan and
Blinnikov (1974). More specifically, for each configuration
we constructed a series of models preserving
the angular momentum distribution $j(m)=\omega(\xi)\xi^2$
from the dimensionless mass coordinate inside the
star $m=M(\xi)/\WI{M}{s}$, where $\xi$ is the cylindrical radius
(cylindrical coordinate) and Ms is the total mass of
the star. The following condition should be fulfilled
for the configuration to be stable:
\begin{equation}
\left(\frac{\partial M}{\partial\WI{\rho}{c}}\right)_{\!\! j(m)}\! >0, \label{dMdrho}
\end{equation}
where $\WI{\rho}{c}$ is the central density and the derivative
in (\ref{dMdrho}) is taken along the series of models. All of
the stellar configurations considered turned out to be
stable. This is not surprising, because the hydrodynamic
stability of LMNSs is determined mainly by
the properties of their massive cores. However, even
the critical rotation for their envelopes turns out to be
extremely weak for the central stellar regions. The
mass outflow from the equator begins much earlier
than the hydrodynamic stability is lost. However, the
situation can change under differential rotation, in the
case where the core rotates much faster than the envelope
(see also Appendix A). This question requires
a special consideration; we are planning to carry out
an appropriate study in the immediate future.

\begin{figure}[htb]
\epsfxsize=16cm \hspace{-2cm}\center\epsffile{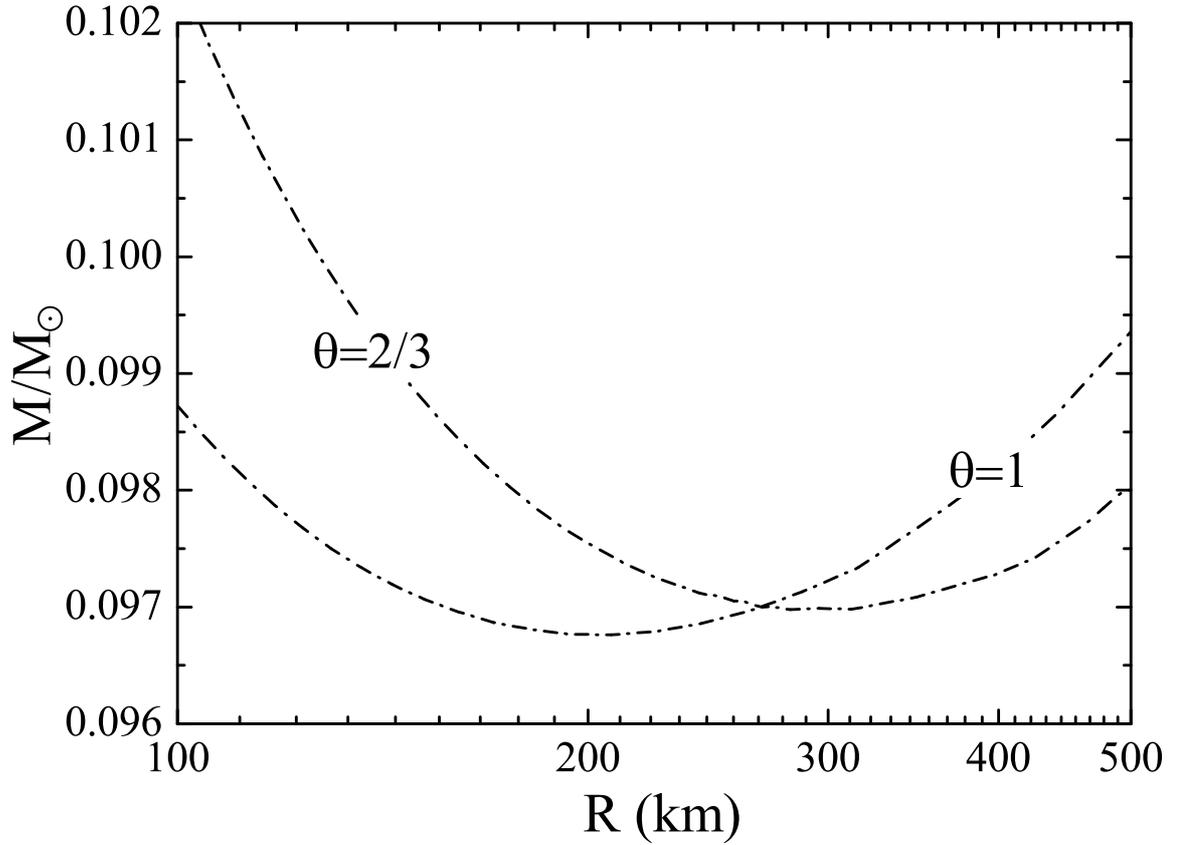}
\caption{\rm Region of the NS mass--radius diagram near the minimum. The lines of non-rotating, $\theta=1$, and critically rotating,
$\theta=2/3$, configurations are shown.}
\label{Pix-Mmin}
\end{figure}
The second question that should be discussed are
the properties of LMNSs with rotation in the region
of the minimum mass. Figure~\ref{Pix-Mmin} shows the lower part
of the LMNS mass–radius diagram on an enraged
scale. Two limiting curves are plotted: the $\theta=1$ curve
without rotation and the $\theta=2/3$ curve with critical
(in the sense of an outflow) rotation. This region of the
diagram is particularly interesting, because here the
LMNS loses its stability and experiences an explosive
expansion (Blinnikov et al. 1990). Accordingly, rotation
can affect the point of stability loss and, hence,
the explosion parameters. Figure~\ref{Pix-Mmin} shows that the
minimum mass changes (increases) insignificantly
even for the case of critical rotation. The radius of the
last stable non-rotating configuration is slightly more
than 200~km (see Table 1), while for the rotating one
it is $\sim$300~km. Given that $\theta$ is $2/3$ here, this increase
in the radius is explained simply by the deformation of
the light extended envelope. It can be concluded that
rigid rotation affects insignificantly the properties of
the lower NS mass limit and, hence, the parameters
of the explosion following the loss of LMNS stability.

\section*{CONCLUSIONS}
\noindent Let us list and discuss the main results of this
paper. First, we showed that all of the admissible
states of rigidly rotating LMNSs occupy a comparatively
narrow region (Fig.~\ref{Pix-Main}) bounded by the curves
with constant oblateness on the mass–radius diagram:
$\theta=1$ (non-rotating configurations) and $\theta=2/3$. The reason why the latter value was singled
out is explained in Appendix A. All configurations
in the region under consideration (``horn'') turn out
to be hydrodynamically stable, while a mass outflow
from the equator begins in the star as the $\theta=2/3$ line
is approached (Fig.~\ref{Pix-Levels}, which shows the structure of
several critical configurations). Second, we showed
(Fig.~\ref{Pix-Mmin}) that rigid rotation affects weakly the minimum
NS mass. This conclusion seems important
from the viewpoint of studying the conditions accompanying
the loss of LMNS stability and the ensuing
processes.

However, we deem the derived evolution paths (the
solid curves in Fig.~\ref{Pix-Main}) corresponding to the loss of
mass by the star from the equator to be the most
important result. This course of events seems most
probable in the case of mass transfer in a close NS binary.
Two NSs approach each other due to the loss of
angular momentum through gravitational radiation.
Let us first consider the case of a non-rotating LMNS:
during the approach it is the first to fill its Roche
lobe. As soon as part of its mass is transferred to
the more massive companion, the separation between
the stars increases, because the binary becomes more
asymmetric. However, the LMNS radius also increases
(Fig.~\ref{Pix-Main}). Hence, two case can be realized:
at a sufficiently large LMNS mass this increase in its
radius is not enough for the accretion to continue, and
the binary should again lose its angular momentum
due gravitational radiation before a new mass transfer.
Thus, the evolution of the binary at this stage will be
governed precisely by the (slow) radiation of gravitational
waves. If, alternatively, the LMNS mass is
sufficiently small, then its radius increases faster than
the components fly apart, and the binary will evolve
on the hydrodynamic (fast) time scale.

Let us now consider the case of mass transfer in a
binary with a LMNS that has a fairly strong rotation.
The LMNS evolution will take the path indicated in
Fig.~\ref{Pix-Main} by the solid lines in the direction of the arrows.
Thus, its radius will drop until the LMNS loses
almost all of its angular momentum! This means
that during all of this period the binary evolution rate
will be determined by the slow energy loss through
gravitational radiation. Only having gotten rid of the
rotation almost completely does the LMNS track turn
and asymptotically approach the $\theta=1$ line of non-rotating
configurations. Subsequently, its evolution
differs little from that for a star without rotation. Thus,
the LMNS spin can only lengthen the evolution time
of a NS binary system, but the LMNS approaches the
minimum mass virtually non-rotating and, hence, the
parameters of its explosion, which marks the loss of
stability by it, will be the same as those in the absence
of rotation. This conclusion seems particularly important
to us.

The work of A.V. Yudin and S.I. Blinnikov was supported
by the Russian Foundation for Basic Research
(project no. 18-29-021019 mk).
\\

\appendix{APPENDIX A: MAXIMUM OBLATENESS IN THE ROCHE
MODEL}

\noindent Let us describe the reason why the limiting LMNS
oblateness is close to $2/3$ (see also Krat 1950). Consider
a rotating axisymmetric star. The equilibrium
equations will be written as (Tassoul 1978)
\begin{align}
\frac{1}{\rho}\frac{dP}{d\xi}&=-\frac{d\WI{\varphi}{G}}{d\xi}+\omega^2 \xi,\label{dP1}\\
\frac{1}{\rho}\frac{dP}{d z}&=-\frac{d\WI{\varphi}{G}}{d z}.\label{dP2}
\end{align}
Here, $P$ is the pressure, $\rho$ is the matter density, $\WI{\varphi}{G}$
is the gravitational potential, and $\omega$ is the angular
velocity of rotation. Equation (\ref{dP1}) describes the
equilibrium of matter in a plane perpendicular to the
rotation axis, $\xi=\sqrt{x^2+y^2}$ is the cylindrical radius.
In the analogous Eq. (\ref{dP2}) $z$ is the coordinate along
the rotation axis. Suppose that the equation of state
is barotropic, i.e., $P=P(\rho)$ (this is definitely the case
for a cool NS considered by us). Then, introducing
the enthalpy
\begin{equation}
H(\rho)\equiv\int\frac{dP'}{\rho'},
\end{equation}
and integrating Eq. (\ref{dP2}), we will obtain $H+\WI{\varphi}{G}=F(\xi)$, where $F(\xi)$ is some function. Substituting this
expression into (\ref{dP1}), we will obtain the Poincare
theorem on the constancy of the angular velocity on
cylindrical surfaces coaxial to the rotation, $\omega=\omega(\xi)$,
and the explicit form of the function $F(\xi)$ itself:
\begin{equation}
F(\xi)=H+\WI{\varphi}{G}=\int \omega^2\xi d\xi+\mathrm{const}.\label{H_plus_phi}
\end{equation}
In the Roche model the entire mass is concentrated
at the center, and the gravitational potential can be
specified in explicit form: $\WI{\varphi}{G}=-\frac{G M}{r}$. LMNSs have
a tiny dense core, where almost all of their mass is
concentrated, and an extended tenuous envelope (see,
e.g., Fig.~\ref{Pix-Levels}). Therefore, the Roche model must well
describe our case, at least to a first approximation. Let
the equatorial and polar radii of the star be $\WI{R}{e}$ and $\WI{R}{p}$,
respectively. From (\ref{H_plus_phi}) we then obtain
\begin{equation}
\int\limits_0^{\WI{R}{e}} \omega^2\xi d\xi=\WI{\varphi}{G}(\WI{R}{e}){-}\WI{\varphi}{G}(\WI{R}{p})=
GM\left(\frac{1}{\WI{R}{p}}-\frac{1}{\WI{R}{e}}\right).\label{w_int}
\end{equation}
Under critical rotation a mass outflow from the equator
begins in the star, i.e., the following condition is
fulfilled:
\begin{equation}
\omega^{2}(\WI{R}{e})\WI{R}{e}=\frac{GM}{\WI{R}{e}^2}.
\end{equation}
Let us introduce the dimensionless quantities
$\varpi\equiv\omega(\xi)/\omega(\WI{R}{e})$ and $\zeta=\xi/\WI{R}{e}$. Relation (\ref{w_int}) will then
be written as
\begin{equation}
\frac{\WI{R}{e}}{\WI{R}{p}}=1+\int\limits_0^{1}\varpi^2(\zeta)\zeta d\zeta,\label{rAB}
\end{equation}
with the constraints
\begin{equation}
\varpi(1)=1, \quad \frac{d\varpi(\zeta)\zeta^2}{d\zeta}\geq 0.
\end{equation}
The latter Solberg–Hoiland condition provides the
rotation stability (for more details, see Tassoul 1978).
For rigid rotation
 $\varpi=\mathrm{const}$, and we get $\WI{R}{e}/\WI{R}{p}=3/2$, which is what was required to prove. The widely
used rotation law, which, given the normalization,
is written as
$\varpi(\zeta)=(1+\alpha)/(1+\alpha\zeta^2)$, where $\alpha$ is
some positive constant, leads to the ratio $\WI{R}{e}/\WI{R}{p}=(3+\alpha)/2$. Thus, generally speaking, differentially rotating
LMNS can be more oblate, up to $\theta=\WI{R}{p}/\WI{R}{e}<2/3$.

\vspace{1 cm}

\appendix{APPENDIX B: THE CENTRAL BOUNDARY CONDITION
IN THE ROTAT CODE}

\noindent The boundary conditions correspond to symmetry
at the center: for $r\rightarrow 0$ we have
\begin{equation}
\int\limits_0^\pi\nabla^2 \Phi\sin\vartheta d\vartheta\rightarrow
3\int\limits_0^\pi\frac{\partial^2 \Phi}{\partial  r^2}\sin\vartheta d\vartheta \; ,
\label{intTheta}
\end{equation}
so that
\begin{eqnarray}
     4\pi G \WI{\rho}{c}
     & = & \frac{1}{2} \int\limits_0^\pi\nabla^2 \Phi\sin\vartheta d\vartheta \; \bigg|_{  r=0} \nonumber \\
    & \simeq & \frac{6}{(\Delta  r)^2}
     \sum_k\
         \Bigl\{
             \cos\bigl[\max(0,\Delta\vartheta(k-1/2))\bigr] \nonumber \\
           & - &\cos\bigl[\min(\pi/2,\Delta\vartheta(k+1/2))\bigr]
         \Bigr\}
         ( \Phi_{1,k} - \Phi_{0,k}),
\end{eqnarray}
where we use the notation $\Phi_{i,k}\equiv \Phi(  r_i,\vartheta_k)$, and the
summation over $k$ to take the integral (\ref{intTheta}) is from
the pole to the equator. The coefficient $6$ in the last
equation is derived from the product of $3$ in (\ref{intTheta}) and
$2$ in the approximation of the second derivative with
respect to the radius at the center:
\begin{equation}
\frac{\partial^2 \Phi}{\partial  r^2} = 2 \frac{\Phi_{1,k}-\Phi_{0,k}}{(\Delta r)^2} .
\end{equation}

\pagebreak

\end{document}